\documentclass[11pt]{article}
\usepackage{geometry}
\usepackage{makeidx}
\usepackage[T1]{fontenc}
\usepackage[dvipsnames,svgnames,table]{xcolor}
\usepackage{epstopdf}
\usepackage{epsfig}
\usepackage{textcomp}
\usepackage{hyperref}
\usepackage{amsmath}
\usepackage{amssymb}
\usepackage{multirow}
\usepackage{multicol}
\usepackage{booktabs}
\usepackage{lastpage}
\usepackage{hyperref} 
\usepackage{graphicx}
\usepackage{enumerate}
\usepackage{threeparttable}
\usepackage{float}
\usepackage{array}
\usepackage{cite}
\usepackage{color,soul}
\makeatletter
\renewcommand\section{\@startsection{section}{1}{\z@}%
  {-2.5ex \@plus -1ex \@minus -.2ex}%
  {2.3ex \@plus.2ex}%
  {\normalfont\large\bfseries}}
\renewcommand\subsection{\@startsection{subsection}{1}{\z@}%
  {-2.5ex \@plus -1ex \@minus -.2ex}%
  {2.3ex \@plus.2ex}%
  {\small\bfseries}}
\begin{document}
\title{\vskip -2.5 cm\textbf{Non-destructive interrogation of nuclear waste barrels through muon tomography: A Monte Carlo study based on dual-parameter analysis via GEANT4 simulations}}
\medskip
\author{\small A. Ilker Topuz$^{1,2}$, Madis Kiisk$^{1,3}$, Andrea Giammanco$^{2}$}
\medskip
\date{\small$^1$Institute of Physics, University of Tartu, W. Ostwaldi 1, 50411, Tartu, Estonia\\
$^2$Centre for Cosmology, Particle Physics and Phenomenology, Universit\'e catholique de Louvain, Chemin du Cyclotron 2, B-1348 Louvain-la-Neuve, Belgium\\
$^3$GScan OU, Maealuse 2/1, 12618 Tallinn, Estonia}
\maketitle
\begin{abstract}
The structural characterization of the nuclear materials constitutes an indispensable aspect that necessitates a careful transportation, a limited interaction, and under certain circumstances an on-site investigation for the nuclear fields including but not limited to nuclear waste management, nuclear forensics, and nuclear proliferation. To attain this purpose, among the promising non-destructive/non-hazardous techniques that are performed for the interrogation of the nuclear materials is the muon tomography where the target materials are discriminated by the interplay between the atomic number, the material density, and the material thickness on the basis of the scattering angle and the absorption in the course of the muon propagation within the target volume. In this study, we employ the Monte Carlo simulations by using the GEANT4 code to demonstrate the capability of muon tomography based on the dual-parameter analysis in the examination of the nuclear waste barrels. Our current hodoscope setup consists of three top and three bottom plastic scintillators made of polyvinyl toluene with the thickness of 0.4 cm, and the composite target material is a cylindrical nuclear waste drum with the height of 96 cm and the radius of 29.6 cm where the outermost layer is stainless steel with the lateral thickness of 3.2 cm and the filling material is ordinary concrete that encapsulates the nuclear materials of dimensions 20$\times$20$\times$20 cm$^{3}$. By simulating with a narrow planar muon beam of 1$\times$1 cm$^{2}$ over the uniform energy interval between 0.1 and 8 GeV, we determine the variation of the average scattering angle together with the standard deviation by utilizing a 0.5-GeV bin length, the counts of the scattering angle by using a 1-mrad step, and the number of the absorption events for the five prevalent nuclear materials starting from cobalt and ending in plutonium. Via the duo-parametric analysis that is founded on the scattering angle as well as the absorption in the present study, we show that the presence of the nuclear materials in the waste barrels is numerically visible in comparison with the concrete-filled waste drum without any nuclear material, and the muon tomography is capable of distinguishing these nuclear materials by coupling the information about the scattering angle and the number of absorption in the cases where one of these two parameters yields strong similarity for certain nuclear materials. 
\end{abstract}
\section{Introduction}
Identification of the radioactive waste forms in the nuclear waste drums is a legislative process that is administered by the competent local authorities in accordance with the standards defined by International Atomic Energy Agency (IAEA)~\cite{international2009iaea, international2011iaea}. By reminding the present generation of the radioactive waste due to the existing radioactive sources in various fields such as energy, medicine, and mining in addition to the old barrels originated in the past practices~\cite{yim2022addressing}, the formal characterization of the nuclear waste barrels de facto requires particular attention as well as ad hoc treatment.\\
While several different techniques based on gamma-rays and neutrons have been already exercised in order to examine the nuclear waste drums~\cite{international2007iaea}, the muon scattering tomography~\cite{procureur2018muon, bonechi2020atmospheric}, where the target materials, i.e. the volume-of-interest (VOI), are discriminated by tracking the muon life cycle through the utilization of the cosmic-ray muons, is also marked in a notable number of studies~\cite{mahon2013prototype, clarkson2014geant4, clarkson2015characterising, thomay2016passive,frazao2019high, topuz2022chem} as a promising method by highlighting its titles such as non-destructive, non-harmful, and portable. Essentially, the basic postulate of the muon scattering tomography underlines the angular deviation of the propagating muons from the initial trajectory principally depending on the atomic number, the density, and the thickness of the target material, and this angular deflection is conventionally measured by computing the scattering angle. Along with the muon deviation due to the VOI, the tomographic setups based on the muon scattering also impart the muon absorption within the VOI, which might be utilized as a complementary characteristic parameter for the purpose of the material classification.\\
In this study, we computationally explore the nuclear waste drums containing a certain amount of bulky radioactive volume~\cite{frazao2019high} by aiming at revealing the quantitative information via the dual combination of the muon scattering angle and the muon absorption. We employ the Monte Carlo simulations by using the GEANT4 code~\cite{agostinelli2003geant4} over our tomographic system~\cite{georgadze2021method} that consists of three plastic scintillators made out of polyvinyl toluene with a thickness of 0.4 cm as well as an accuracy of 1 mrad in both the top section and the bottom section~\cite{thomay2016passive, frazao2019high} and we follow an experimentally repeatable procedure founded on the hit locations in the detector layers. This study is organized as follows. In section~\ref{sec:Definition of characteristic parameters}, we express the characteristic parameters, i.e. the scattering angle and the relative absorption rate, for the discrimination of the nuclear waste barrels including different types of nuclear materials. While we present the hodoscope layout as well as the simulation properties in section~\ref{sec:Hodoscope scheme and simulation properties}, the simulations results are exhibited by using both quantitative and qualitative formats in section~\ref{sec:Simulation outcomes}, and we state our concluding remarks in section~\ref{sec:Conclusion}. 
\label{sec:introduction}
\section{Definition of characteristic parameters}
\label{sec:Definition of characteristic parameters}
\subsection{Average scattering angle and standard deviation}
\begin{figure}[H]
\begin{center}
\includegraphics[width=7.5cm]{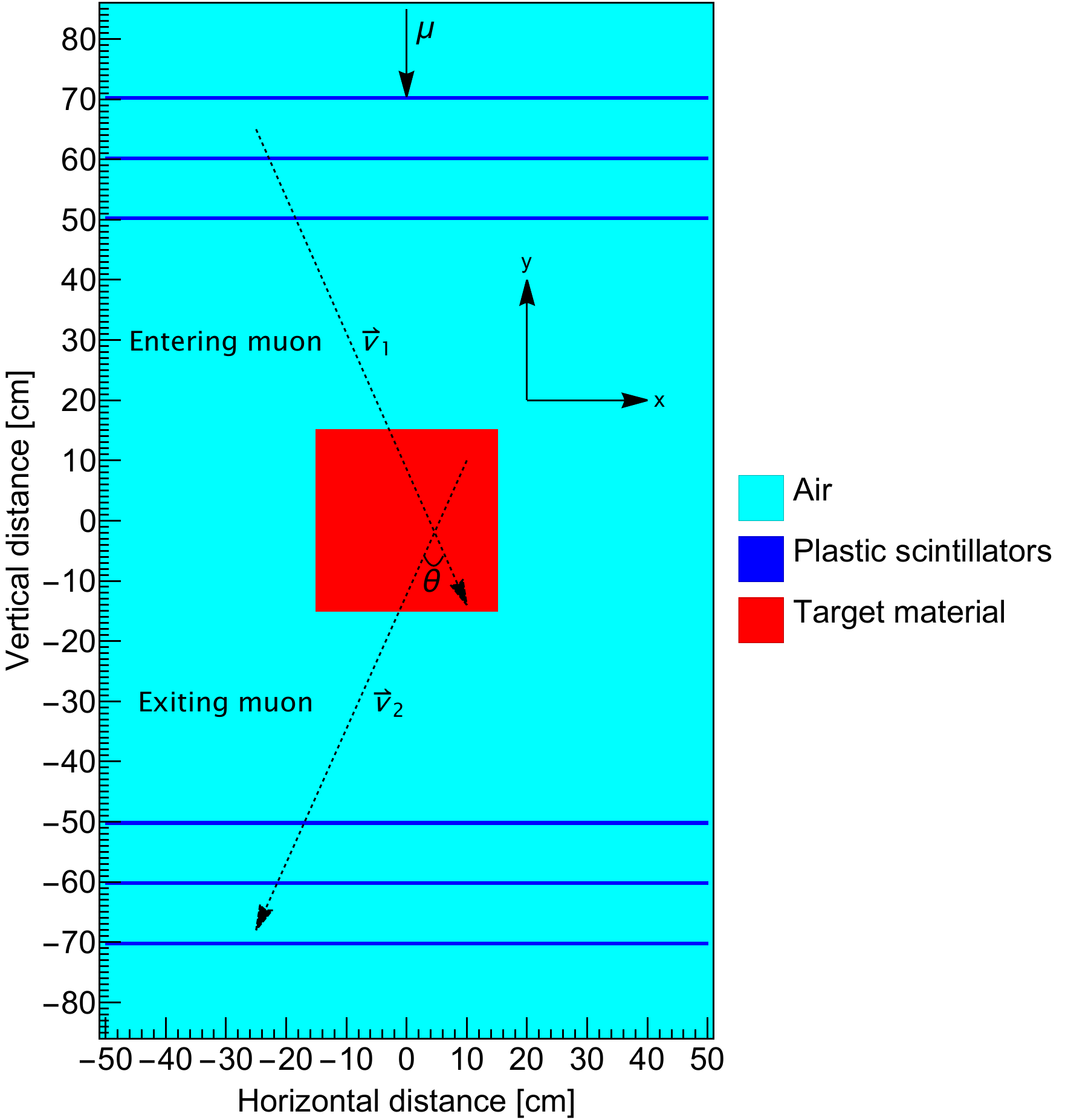}
\caption{Definition of scattering angle denoted by $\theta$ according to the hit locations in the detector layers.}
\label{Scattering angle}
\end{center}
\end{figure}

In the current study, the scattering angle of a muon means the three-dimensional positive angular difference between the direction of the entering muon through the VOI and the direction of the same exiting muon from the same VOI, and this angular deviation is caused by the interactions that stochastically occur between the propagating muons and the VOI. As described in Fig.~\ref{Scattering angle}, the computation of the scattering angle requires the construction of two independent vectors by utilizing exactly four muon hit locations in the detector layers where the first vector is the difference between the hits locations in the second top detector layer and the third top detector layer, while the subtraction of the hit position in the first bottom plastic scintillator from the hit position in the second bottom plastic scintillator yields the latter vector.
The definition of these two vectors brings forth the scattering angle denoted by $\theta$, and the scattering angle of a muon crossing the VOI is obtained by using these two vectors as follows~\cite{carlisle2012multiple, nugent2017multiple, poulson2019application}
\begin{equation}
\theta=\arccos\left (\frac{\vec{v}_{1} \cdot \vec{v}_{2}}{\left|v_{1}\right|\left|v_{2}\right|}\right)
\end{equation}
Since a substantial number of muons reach the VOI, the average profile of the scattering angle at a certain energy is quantified by averaging the previously determined scattering angles over $N$ number of the non-absorbed/non-decayed muons as written in
\begin{equation}
\bar{\theta}=\frac{1}{N}\sum_{i=1}^{N}\theta_{i}
\end{equation}
where its standard deviation is 
\begin{equation}
\delta\theta=\sqrt{\frac{1}{N}\sum_{i=1}^{N}(\theta_{i}-\bar{\theta})^{2}}
\end{equation}
\subsection{Relative absorption rate}
In the course of the muon penetration through the VOI, its kinetic energy is deducted by multiple mechanisms~\cite{groom2001muon}, and this collective slowing-down is implicitly contingent on the intrinsic properties of the VOI as well as the thickness of the VOI. Especially in the case of the relatively low-energetic muons, the energy loss due to the VOI might lead to either the zero-energy value or the quasi-zero-energy level that commonly results in the capture of the corresponding muon at rest. In the GEANT4 nomenclature for a negative muon denoted by \textmu$^{-}$, this process is entitled "muMinusCaptureAtRest", and it might support the material characterization under certain circumstances. Hence, in the present study, we also track the number of the \textmu$^{-}$ captures at rest within the VOI and we define a relative ratio called relative absorption rate (RAR) between the absorbed muons and the generated muons as expressed in
\begin{equation}
\rm RAR=\frac{\rm Absorption}{\rm Generation}=\frac{\mbox{\# of muMinusCaptureAtRest}}{\mbox{\# of \textmu$^{-}$}}
\label{RAR}
\end{equation}
Besides the material properties, since the muon absorption is also dependent on the muon energy spectrum regarding the energy cut-off and the population size of the potentially absorbable muons, the absorption rate in Eq.~(\ref{RAR}) is axiomatically relative.\\
Even in the case of a fair energy cut-off, the non-absorbed muons leaving the VOI might be still subject to the capture at rest in either the surrounding medium or the bottom detector layers in accordance with their final energies, thus we further track the absorption events that occur outside the VOI. 
\section{Hodoscope scheme and simulation properties}
\label{sec:Hodoscope scheme and simulation properties}
\begin{figure}[H]
\begin{center}
\includegraphics[width=8cm]{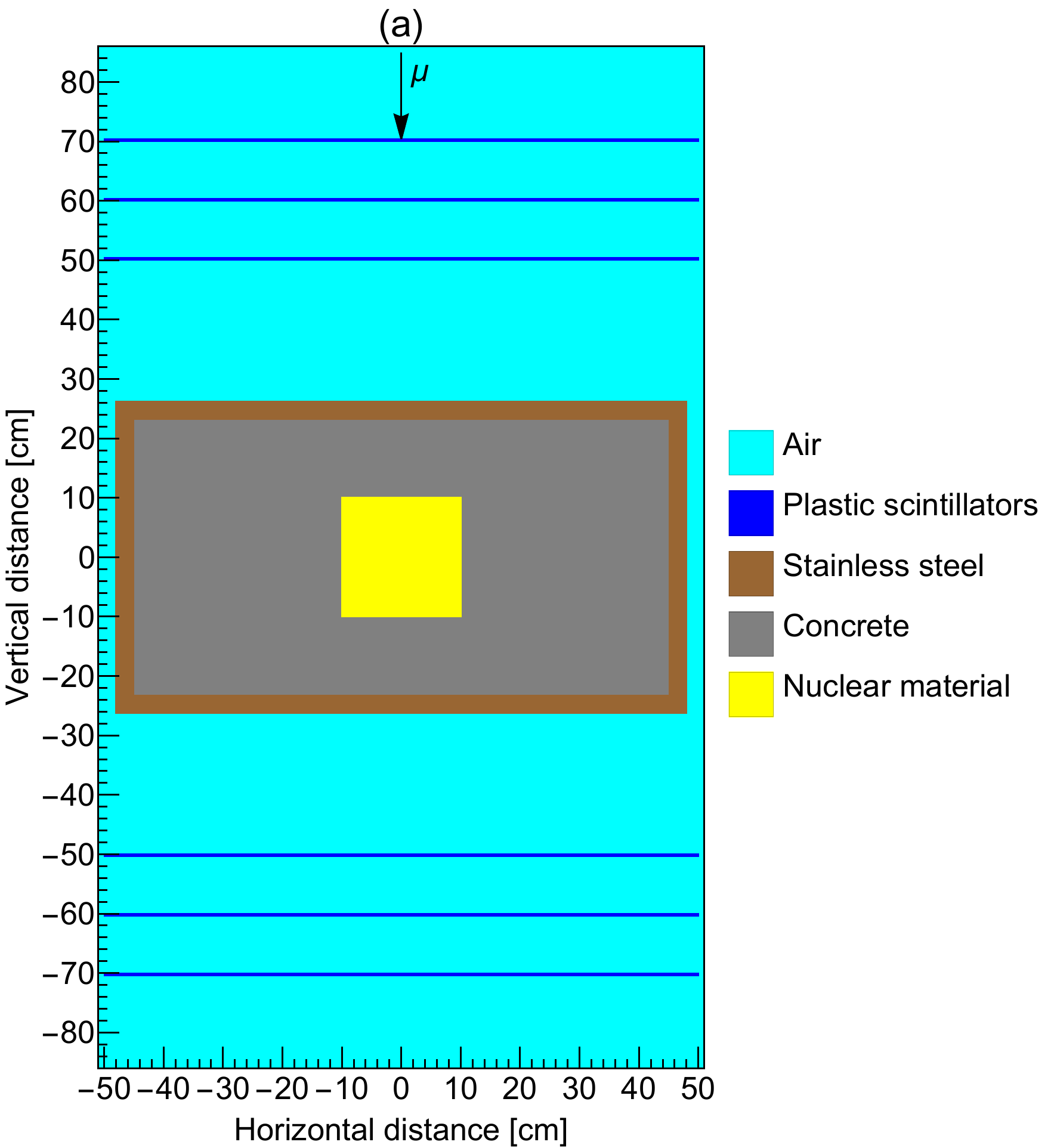}
\hskip 2cm
\includegraphics[width=6.25cm]{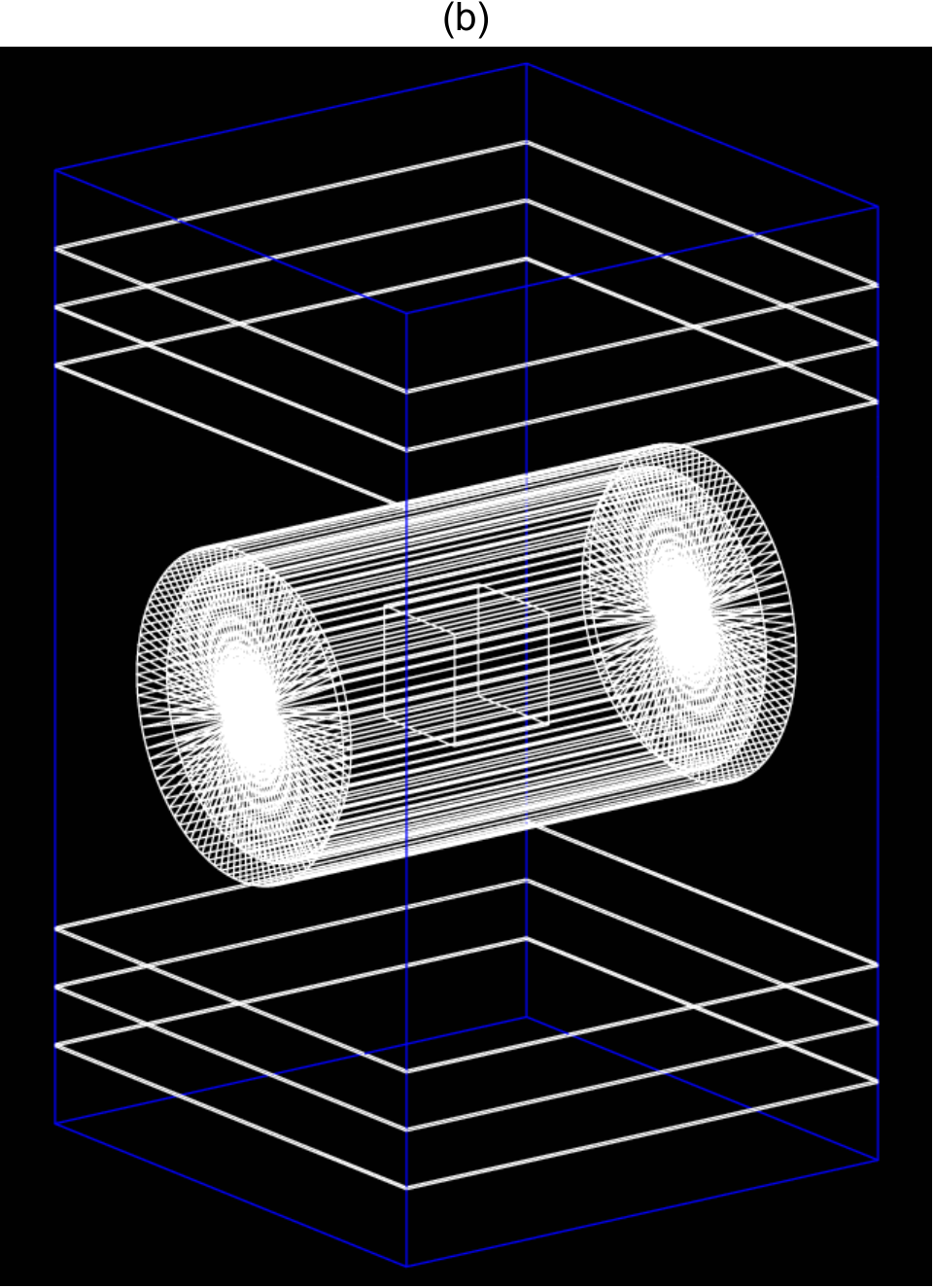}
\caption{Illustration of simulation components (a) layout of the nuclear waste barrel within the tomographic system and (b) reproduced geometry in GEANT4.}
\label{Waste_barrel}
\end{center}
\end{figure}

Heretofore, we have briefly described the dual-parametric approach based on the muon scattering angle and the muon absorption. To perform the aforementioned analysis, the geometrical scheme is depicted in Fig.~\ref{Waste_barrel}, and it is shown that the plastic scintillators are separated by a distance of 10 cm, whereas the distance between these two hodoscope sections is 100 cm. Furthermore, the dimensions of the detector layers are $100\times0.4\times100$ cm$^{3}$. Concerning the nuclear waste drum, the VOI is held at the center of the tomographic system. Regarding the components of the nuclear waste barrel, the outermost layer is defined as a cylinder manufactured from stainless steel layer, the height of which is 96 cm, and the thickness of which is 3.2 cm. The filling material is the cylindrical ordinary concrete slab with the height of 88 cm as well as the radius of 26.2 cm, while the nuclear material placed at the middle of the concrete padding is a cubic solid box of $20\times20\times20$ cm$^{3}$.
\vskip -0.5cm
By fulfilling the geometrical properties of the tomographic setup as well as the regular nuclear waste drum, we conduct the Monte Carlo simulations via the GEANT4 code in order to register the hit positions in the plastic scintillators. The simulation parameters are listed in Table~\ref{Simulation properties}, and the dimension of the simulation box is $100\times170\times100$ cm$^{3}$ where the Cartesian components are situated symmetrically in the interval of (-50 cm, 50 cm), (-85 cm, 85 cm), and (-50 cm, 50 cm), respectively as indicated in Fig.~\ref{Waste_barrel}(a). Into the bargain, we exhibit the reproduced geometry in GEANT4 as displayed in Fig.~\ref{Waste_barrel}(b). We use a narrow planar multi-energetic mono-directional beam that is generated at ([-0.5, 0.5] cm, 85 cm, [-0.5, 0.5] cm) via G4ParticleGun, and the generated muons are propagating in the vertically downward direction as shown by the black arrow in Fig.~\ref{Waste_barrel}(a), i.e. from the top edge of the simulation box through the bottom edge.
\vskip -0.5cm
\begin{table}[H]
\begin{center}
\caption{Simulation properties.}
\begin{footnotesize}
\begin{tabular}{cc}
\toprule
\toprule
Particle & $\mbox{\textmu}^{-}$\\
Beam direction & Vertical\\
Momentum direction & (0, -1, 0)\\
Source geometry & Planar\\
Initial position (cm) & ([-0.5, 0.5], 85, [-0.5, 0.5])\\
Number of particles & 10$^{5}$\\
Energy interval (GeV) & [0, 8]\\
Energy cut-off (GeV) & 0.1\\
Bin step length (GeV) & 0.5\\
Energy distribution & Uniform\\
Material database & G4/NIST\\
Reference physics list & FTFP$\_$BERT\\
\bottomrule
\bottomrule
\label{Simulation properties}
\end{tabular}
\end{footnotesize}
\end{center}
\end{table}
\vskip -1cm
A uniform energy distribution lying on the interval between 0 and 8 GeV with the energy cut-off of 0.1 GeV, which is selected to minimize the probability of the muon absorption in the top detector layers as well as to maximize the encounter between the incoming muons and the VOI, is utilized by recalling the numerical advantages~\cite{anghel2015plastic}. The total number of the generated \textmu$^{-}$ is $10^{5}$ in every simulation. All the materials in the simulation geometry are defined in agreement with the GEANT4/NIST material database, and FTFP$\_$BERT is the reference physics list used in the present study.\\
The muon tracking is maintained by G4Step, and the registered hit locations are post-processed by the aid of a Python script where the scattering angle is first calculated for every single non-absorbed/non-decayed muon, then the uniform energy spectrum bounded by 0 and 8 GeV is partitioned into 16 bins by marching with a step of 0.5 GeV, and each obtained energy bin is labeled with the central point in the energy sub-interval. Consequently, the obtained scattering angles are averaged for the associated energy bins. In the case of the muon capture at rest, the in-target absorption is acquired by directly probing the VOI, which also means that the events called "muMinusCaptureAtRest" are recorded during the muon propagation within the VOI.
\section{Simulation outcomes}
\label{sec:Simulation outcomes}
\begin{figure}[H]
\setlength{\belowcaptionskip}{-4ex} 
\begin{center}
\includegraphics[width=9.25cm]{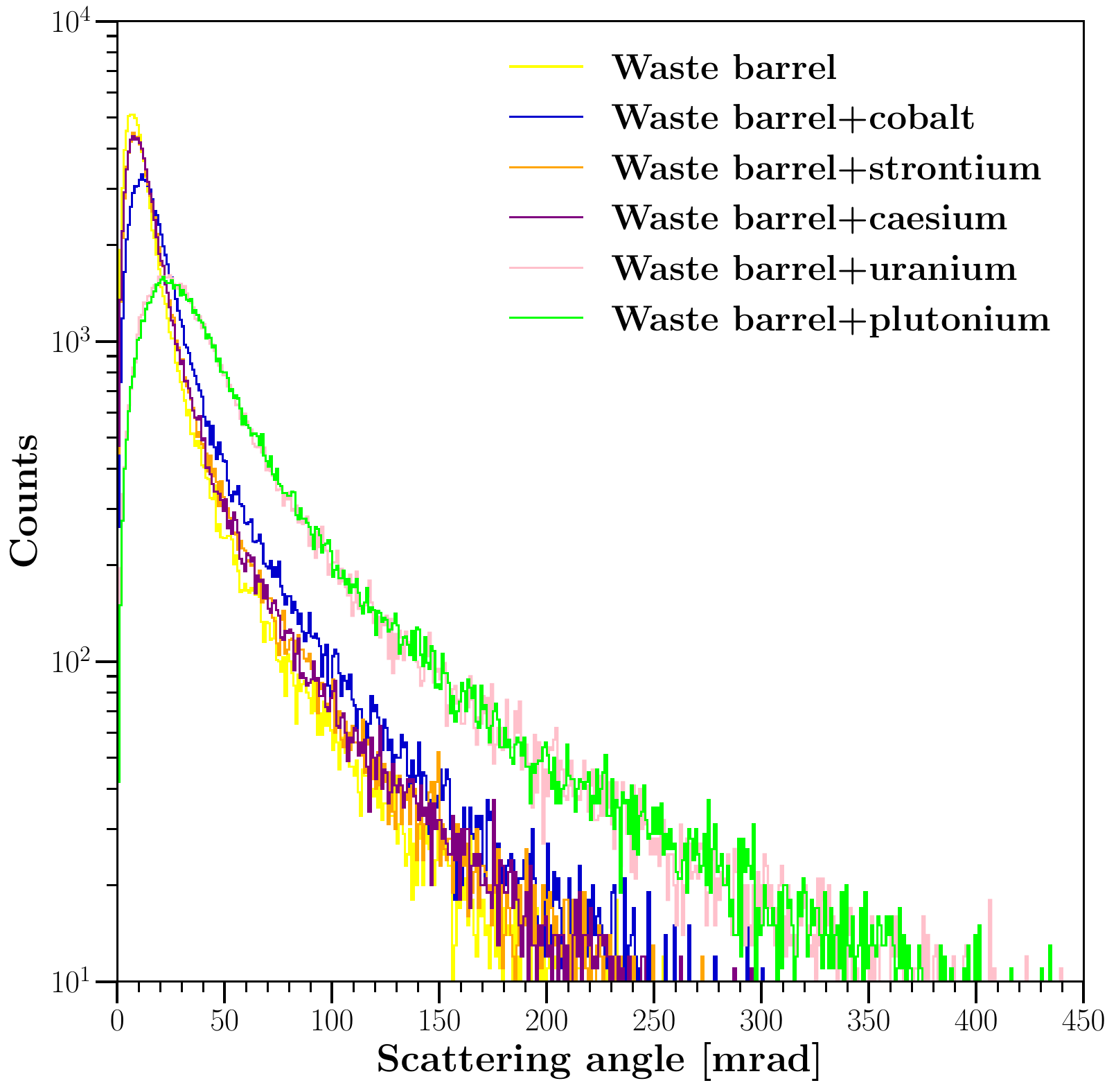}
\caption{Distribution of the scattering angles for the nuclear waste barrels with a step length of 1 mrad over the energy interval between 0.1 and 8 GeV.}
\label{Distribution}
\end{center}
\end{figure}

To test the feasibility of the dual-parametric methodology by using the above-mentioned simulation setup, we select a list of nuclear materials composed of caesium, strontium, cobalt, uranium, and plutonium. Accompanying the nuclear waste barrels that contain these bulky nuclear materials, we also define an ordinary waste drum denoted by waste barrel or WB that only consists of stainless steel and concrete for the sake of comparison. On the first basis, we first determine the scattering angle distribution by using a 1-mrad step length, and Fig.~\ref{Distribution} depicts the distribution of the scattering angles for the nuclear waste barrels over the energy interval between 0.1 and 8 GeV. 
\begin{table*}[!ht]
\begin{footnotesize}
\begin{center}
\caption{Average scattering angles of the nuclear waste barrels and their corresponding standard deviations over the energy interval between 0.1 and 8 GeV.}
\resizebox{\textwidth}{!}{\begin{tabular}{*7c}
\toprule
\toprule
$\bar{E}$ [GeV] & $\bar{\theta}_{\rm WB}\pm\delta\theta$ [mrad] & $\bar{\theta}_{\rm WB+Co}\pm\delta\theta$ [mrad] & $\bar{\theta}_{\rm WB+Sr} \pm\delta\theta$ [mrad] & $\bar{\theta}_{\rm WB+Cs} \pm\delta\theta$ [mrad] & $\bar{\theta}_{\rm WB+U} \pm\delta\theta$ [mrad] & $\bar{\theta}_{\rm WB+Pu} \pm\delta\theta$ [mrad] \\
\midrule
0.25	& 182.389$\pm$117.135 & - & 201.051$\pm$124.601 &199.948$\pm$120.593&- & -\\							
0.75	& 78.994$\pm$47.122 & 157.400$\pm$105.707& 92.951$\pm$55.837 & 89.627$\pm$53.821& 289.983$\pm$161.253&297.717$\pm$163.556\\
1.25	& 43.522$\pm$24.559 & 75.429$\pm$42.140 & 51.847$\pm$29.057 & 51.126$\pm$28.367& 168.248$\pm$92.783&173.095$\pm$96.537\\
1.75	& 30.718$\pm$17.368 & 50.800$\pm$27.873& 36.381$\pm$20.748 & 35.690$\pm$19.545& 108.194$\pm$58.337&110.977$\pm$60.585\\
2.25	& 23.602$\pm$12.956 & 38.686$\pm$20.690&27.900$\pm$15.105 & 27.553$\pm$14.894& 81.011$\pm$43.143&83.956$\pm$45.269\\
2.75	& 19.446$\pm$10.581 & 31.110$\pm$16.801&22.672$\pm$12.227 & 22.315$\pm$11.966& 63.784$\pm$33.997&66.758$\pm$36.016\\
3.25	& 16.117$\pm$8.957 & 26.122$\pm$14.312 &19.382$\pm$10.527 & 19.236$\pm$10.413& 53.779$\pm$28.704&55.828$\pm$29.753\\
3.75	& 14.108$\pm$7.820 & 22.599$\pm$11.932 &16.366$\pm$8.758 & 16.414$\pm$8.826&46.368$\pm$24.609&47.536$\pm$25.213\\
4.25	& 12.254$\pm$6.802 & 20.111$\pm$10.894 &14.590$\pm$7.974 & 14.473$\pm$7.845&40.141$\pm$21.139&41.817$\pm$22.291\\
4.75	& 10.822$\pm$5.927 & 17.527$\pm$9.548 &13.051$\pm$7.402 & 12.909$\pm$7.035&36.105$\pm$18.756&37.172$\pm$19.537\\
5.25	& 10.014$\pm$5.873 & 15.987$\pm$8.576 &11.667$\pm$6.399 & 11.702$\pm$6.427&32.208$\pm$18.083&33.383$\pm$17.889\\
5.75	& 8.980$\pm$4.926 & 14.466$\pm$7.934&10.649$\pm$5.684 & 10.733$\pm$5.894&29.290$\pm$16.529&30.168$\pm$16.031\\
6.25	& 8.390$\pm$4.524 & 13.295$\pm$7.162&9.805$\pm$5.398 & 9.770$\pm$5.299&26.823$\pm$14.305&27.933$\pm$15.374\\
6.75	& 7.723$\pm$4.286 & 12.335$\pm$6.522 &9.203$\pm$5.396 & 9.115$\pm$4.944&24.759$\pm$12.938&25.357$\pm$13.712\\
7.25	& 7.256$\pm$5.968 & 11.427$\pm$6.169 &8.535$\pm$4.688 & 8.471$\pm$4.631&23.085$\pm$12.056&23.903$\pm$12.730\\
7.75	& 6.789$\pm$5.355 & 10.525$\pm$5.626 &8.037$\pm$5.289 & 7.891$\pm$4.320&21.386$\pm$11.532&22.359$\pm$11.670\\
\bottomrule
\bottomrule
\label{average_scattering_angle and standard deviation}
\end{tabular}}
\end{center}
\end{footnotesize}
\end{table*}
It is demonstrated that both the nuclear waste drums including strontium and caesium exhibit a close trend compared to the waste barrel, while both the nuclear waste barrels encompassing plutonium and uranium yield significantly distinct scattering angle profiles due to their high atomic numbers and the high density values in comparison with the waste barrel as well as the rest of nuclear waste barrels. Thus, regarding the practical efficiency of the material discrimination, the region around uranium along with the trans-uranium elements in the periodic table shows a remarkable advantage contrary to the other materials. It is worth mentioning that, for this specific setup that assumes the bulky radioactive volume, a nuclear waste drum containing cobalt also displays a visibly different distribution.

\begin{figure}[H]
\setlength{\belowcaptionskip}{-4ex} 
\begin{center}
\includegraphics[width=9.25cm]{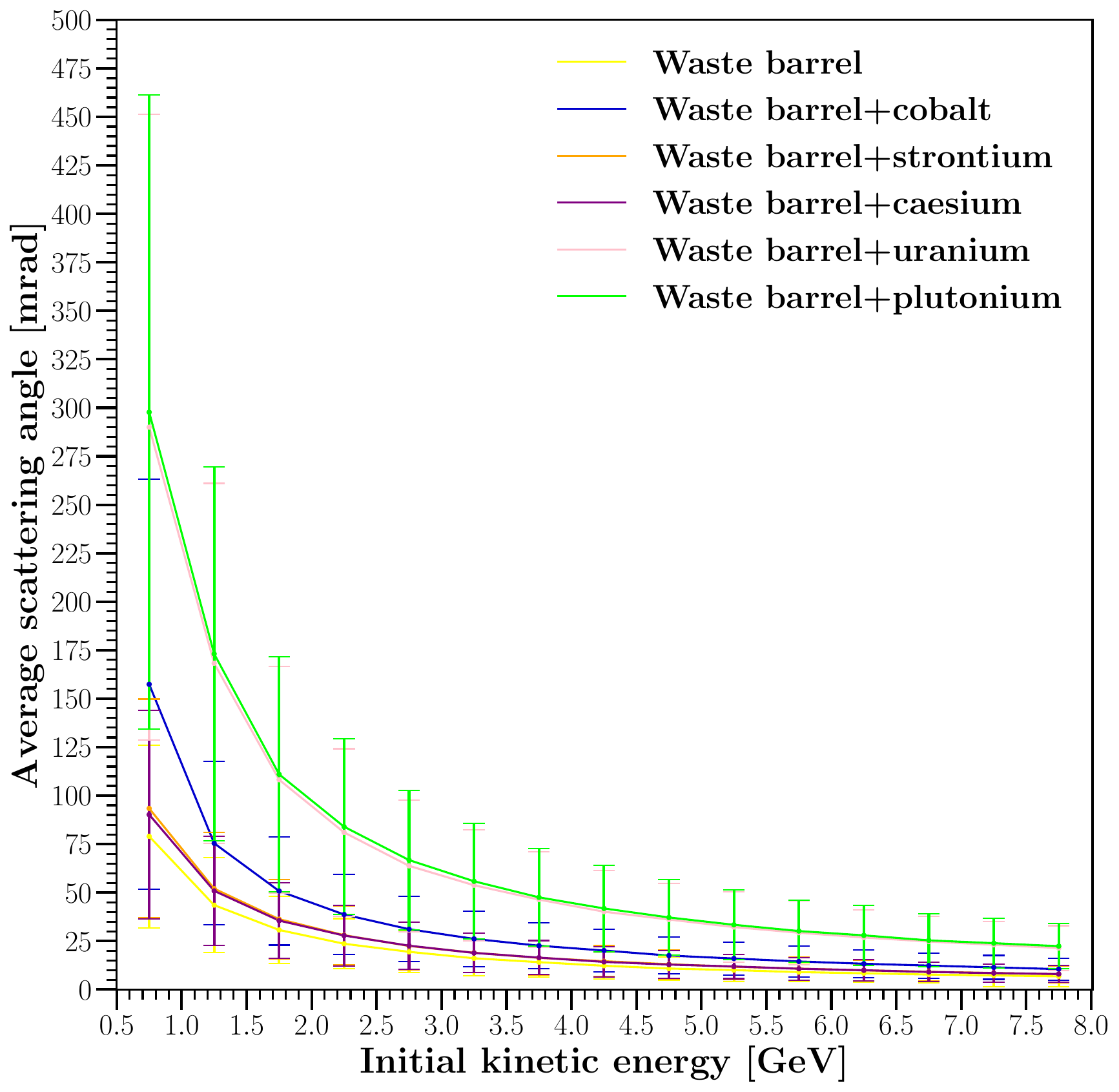}
\caption{Variation of the average scattering angle with respect to the energy bins with a bin length of 0.5 GeV over the energy interval between 0.1 and 8 GeV by recalling the standard deviations listed in Table~\ref{average_scattering_angle and standard deviation}.}
\label{Exponential}
\end{center}
\end{figure}

Whereas the distribution of the scattering angle provides a qualitative profile for the initial evaluation, we calculate the average scattering angle and the corresponding standard deviation for a set of 16 energy bins in order to obtain the quantitative details about the present nuclear waste barrels, and Table~\ref{average_scattering_angle and standard deviation} tabulates the average scattering angles and the standard deviations over the energy interval between 0.1 and 8 GeV. According to Table~\ref{average_scattering_angle and standard deviation}, the nuclear waste drums containing uranium or plutonium generate similar scattering angles, and the nuclear waste barrels including strontium and caesium give rise to the close scattering angles. Although it is partially hard to claim a remarkable difference between the waste barrel and the nuclear waste drums having strontium and caesium by just checking the distribution of the scattering angles in Fig.~\ref{Distribution}, Table~\ref{average_scattering_angle and standard deviation} indicates a slight difference between these cases, thereby providing a challenging possibility for the material identification. By analyzing Table~\ref{average_scattering_angle and standard deviation}, it is also revealed that the average scattering angle exponentially declines with respect to the energy increase as shown in another study with the root-means-square values~\cite{luo2016energy}. The variation of the average scattering angle as a fuction of the kinetic energy is  illustrated in Fig.~\ref{Exponential}, and it is seen that the angular difference between the nuclear waste drums decreases when the initial kinetic energy increases.
\vskip -0.25cm 
A similar consequence from Fig.~\ref{Distribution} might be drawn from Fig.~\ref{Exponential} by stating that the nuclear waste barrel with the bulky cobalt acts as a considerable deflector against the propagating muons following the uranium- and plutonium-containing barrels; however, the deflecting capability of the nuclear waste drum with strontium or caesium is not significantly different from that of the waste barrel considering the present configuration. 

The qualitative information as well as the numerical data already shows that the nuclear waste barrels might be classified according to the scattering angle that is directly dependent on the constituents in the nuclear waste barrels. As a matter of fact, Table~\ref{average_scattering_angle and standard deviation} already implies the second characteristic parameter that might be utilized in order to identify the nuclear waste barrels. When the initial energy bin, which is 0.25 GeV, is examined, it is observed that the nuclear waste drums including uranium, plutonium, and cobalt do not possess any value; on the other hand, the waste barrel and also the nuclear waste drums encompassing strontium and caesium have an average scattering angle at the energy bin of 0.25 GeV. The reason behind this absence might be formulated by either the complete absorption of the penetrating muons within the energy bin of 0.25 GeV in the case of plutonium and uranium or the statistically insufficient number of the surviving muons for the energy bin of 0.25 GeV in the case of cobalt.
\begin{table*}[!ht]
\begin{center}
\begin{minipage}{15cm}
\caption{Number of the absorbed muons within the nuclear waste barrels by using five seed numbers over the energy interval between 0.1 and 8 GeV.}
\begin{tabular}{*8c}
\toprule
\toprule
Material & Seed I & Seed II & Seed III & Seed IV & Seed V & Average & Standard deviation\\
\midrule
WB &2820&2754&2853&2790&2825&2808.4&37.7\\
WB+Co&5048&4932&5178&4991&5142&5058.2&102.4\\
WB+Sr&2697&2628&2702&2670&2715&2682.4&34.5\\
WB+Cs&2397&2353&2423&2408&2445&2405.2&34.3\\
WB+U&7304&7206&7403&7248&7140&7314.2&91.2\\
WB+Pu&7508&7423&7629&7464&7620&7528.8&92.4\\
\bottomrule
\bottomrule
\label{absorption}
\end{tabular}
\end{minipage}
\end{center}
\end{table*}
\vskip -0.75cm
Hence, motivated by this fact, we track the absorbed muons within the nuclear waste barrels by utilizing five seed numbers, and Table~\ref{absorption} lists the number of the muon captures at rest inside the nuclear waste barrels over the energy interval between 0.1 and 8 GeV.
\begin{figure}[H]
\setlength{\belowcaptionskip}{-4ex} 
\begin{center}
\includegraphics[width=9cm]{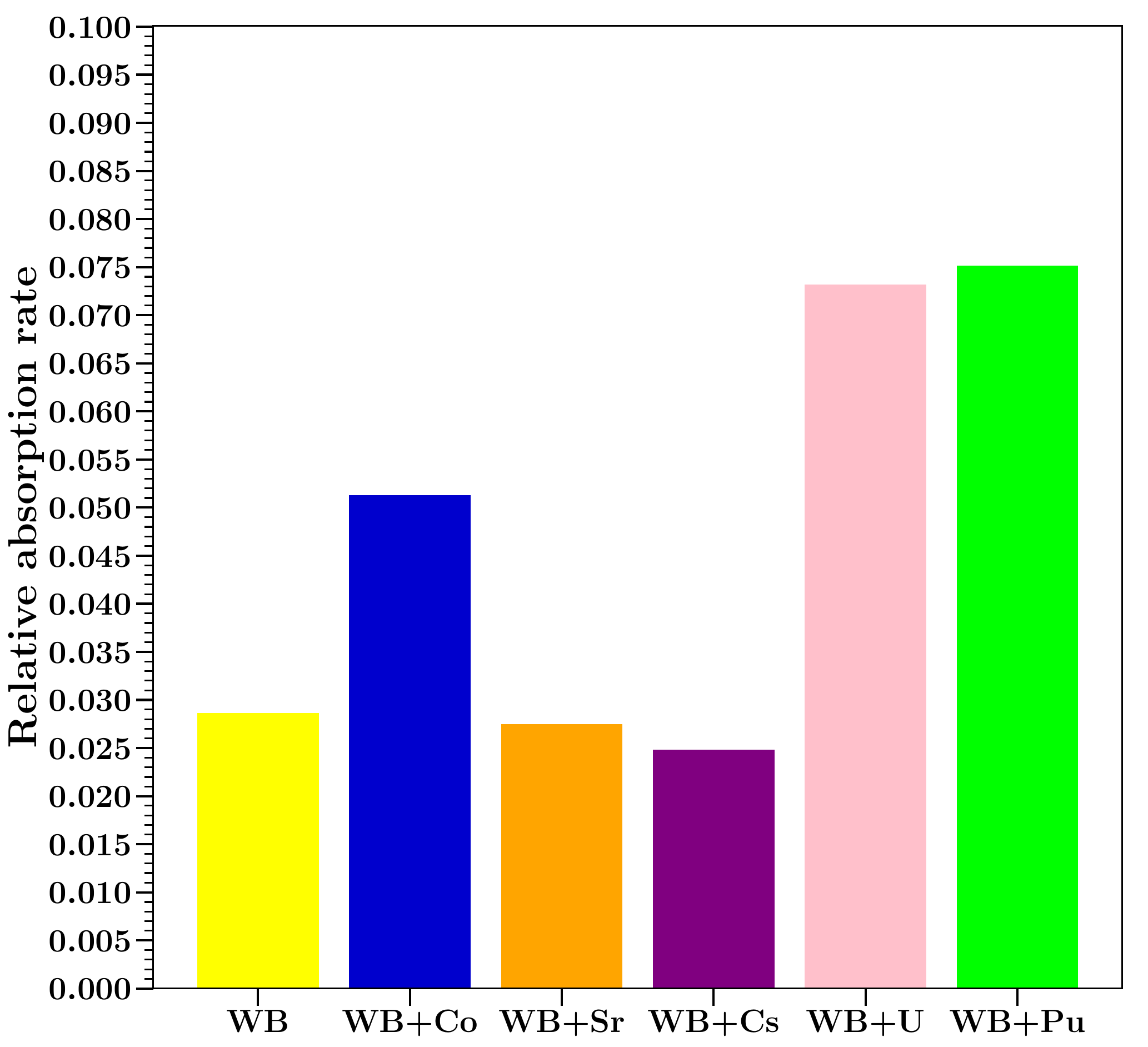}
\caption{Relative absorption rates of the nuclear waste barrels over the energy interval between 0.1 and 8 GeV by reminding the standard deviations tabulated in Table~\ref{absorption}.}
\label{Absrate}
\end{center}
\end{figure}
It is explicitly demonstrated that the nuclear waste drums containing uranium or plutonium yield the highest number of the \textmu$^{-}$ captures at rest as they generate the highest average scattering angles among the current barrels under the investigation; in contrast, the moderating power for the remaining drums except the cobalt case is almost alike. According to our observations, it is worth bearing in mind that a negligible number of muons (about 21 to 40) are subject to the post-target absorption, most of which occur in the bottom detector layers in every case of the present GEANT4 simulations.

In order to compute the RAR as defined in Eq.~(\ref{RAR}), we divide the number of the \textmu$^{-}$ captures at rest within the nuclear waste drums, i.e. the average number of the muon absorption over five seed numbers as indicated in the sixth column of Table~\ref{absorption}, by the total number of the generated muons, which is $10^{5}$ as stated in Table~\ref{Simulation properties}, and Fig.~\ref{Absrate} presents the RAR for the current nuclear waste barrels. However, it is worth noting that a very small portion of the entire muon population usually has the absorption potential, which also means that a statistically reliable absorption dataset undoubtedly requires a long period of muon irradiation.
\vskip -0.75cm
\section{Conclusion}
\label{sec:Conclusion}
In this study, the nuclear waste barrels containing a certain amount of bulky radioactive waste have been quantitatively investigated with regard to the scattering angle as well as the absorption rate by using the GEANT4 simulations for the application in muon scattering tomography. Concerning the scattering angle, we demonstrate that a waste barrel with the bulky cobalt, uranium, and plutonium might be detected by using muon tomography. According to our GEANT4 simulations, we also show that the absorption rate might act as a complementary characteristic parameter in addition to the scattering angle in the case of the mid/high density materials with the condition of the long exposure periods. As a future work, one might utilize these two characteristic parameters in order to train a classifier by aiming at identifying the content of the waste barrels as well as characterizing the performance of this classifier.
\bibliographystyle{elsarticle-num}
\bibliography{dualparameter}
\end{document}